\documentclass[cameraready]{Interspeech}

\title{Two-Stage Adaptation for Non-Normative Speech Recognition: Revisiting Speaker-Independent Initialization for Personalization
\thanks{Under review for Interspeech 2026.}}

\author[affiliation={1}, orcid=0009-0005-4001-8074]{Shan}{Jiang}
\author[affiliation={1}, orcid=0009-0008-0582-3889]{Jiawen}{Qi}
\author[affiliation={2}]{Chuanbing}{Huo}
\author[affiliation={3}, orcid=0009-0000-0876-621X, correspondingauthor]{Yingqiang}{Gao}
\author[affiliation={1}, orcid=0009-0005-9480-6164, correspondingauthor]{Qinyu}{Chen}

\address{
    $^1$ Leiden Institute of Advanced Computer Science, Leiden University, The Netherlands \\
    $^2$ Penobscot Community Health Care, United States \\
    $^3$ Department of Computational Linguistics, University of Zurich, Switzerland
}

\email{q.chen@liacs.leidenuniv.nl, jsmailqwq@gmail.com}

\keywords{automatic speech recognition, non-normative speech, personalization and adaptation.}

\begin{document}
 
\maketitle
\begin{abstract}
    Personalizing automatic speech recognition (ASR) systems for non-normative speech, such as dysarthric and aphasic speech, is challenging.
    While speaker-specific fine-tuning (SS-FT) is widely used, it is typically initialized directly from a generic pre-trained model. Whether speaker-independent adaptation provides a stronger initialization prior under such mismatch remains unclear.
    In this work, we propose a two-stage adaptation framework consisting of speaker-independent fine-tuning (SI-FT) on multi-speaker non-normative data followed by SS-FT, and evaluate it through a controlled comparison with direct SS-FT under identical per-speaker conditions. Experiments on AphasiaBank and UA-Speech with Whisper-Large-v3 and Qwen3-ASR, alongside evaluation on typical-speech datasets TED-LIUM v3 and FLEURS, show that two-stage adaptation consistently improves personalization while maintaining manageable out-of-domain (OOD) trade-offs.
\end{abstract}

\section{Introduction}
Automatic Speech Recognition (ASR) systems pre-trained on large-scale normative speech corpora achieve strong performance on typical speech but degrade significantly when applied to non-normative speech~\cite{de2019impact,shahamiri2021speech}. Speech impairments such as dysarthria (a motor speech disorder affecting articulation) and aphasia (a language production disorder)~\cite{stroke_communication_2024,cleveland_speech_2024} introduce substantial acoustic and lexical variability among individual speakers. These deviations create distribution mismatches that challenge standard ASR foundation models ~\cite{radford2023robust,shi2026qwen3,zhang2023google}.
Pronunciation errors, articulatory instability, disfluencies, and prosodic abnormalities further exacerbate this gap, limiting the deployment of ASR in clinical and assistive settings.
A challenge, therefore, lies in how to effectively adapt large pre-trained ASR models to non-normative speech under limited data.

Recent work has explored various strategies to improve ASR robustness when adapting to non-normative speech.
Input-side approaches focus on reducing acoustic mismatch through speech reconstruction, voice conversion, and dysfluency segmentation and input mixture ratio adjustment~\cite{chen25m_interspeech,elhajal2025unsupervised,ghosh25_interspeech,ASASR2025biocas}. Model adaptation efforts include etiology-specific continued pre-training~\cite{baumann25_interspeech}, self-training on dysarthric speech~\cite{wang2025selftraining}, and parameter-efficient fine-tuning (PEFT) strategies~\cite{tan2025cbawhisper,wagner2025personalized}. Other works improve adaptation stability or deployment through multi-task regularization~\cite{xiong2025mitigating}, federated learning~\cite{zhong2025regularized}, mixture-of-experts (MoE) based speaker adaptation~\cite{hu25d_interspeech}, or LLM-enhanced decoding~\cite{aboeitta2025bridging,qin2025tiny,zhang2025towards}.
While these approaches improve robustness from different perspectives, most focus on modifying model components or training strategies.
Relatively little attention has been paid to the interaction between population-level adaptation and subsequent speaker-level personalization~\cite{green2021automatic}.

In practice, speaker-specific fine-tuning (SS-FT) is often initialized directly from a generic pre-trained foundation ASR model. However, when the target speech distribution deviates substantially from typical training data, this direct personalization may suffer from unstable optimization or inefficient convergence under limited per-speaker data.
This raises a question: When adapting ASR systems to individual non-normative speech, does speaker-independent fine-tuning (SI-FT) serve as a more effective initialization for personalization than direct fine-tuning from a generic pre-trained model?
Intuitively, speaker-independent adaptation on multi-speaker non-normative speech may shift model parameters closer to the target distribution, thereby reducing the distribution gap during personalization and improving sample efficiency. At the same time, such adaptation may introduce trade-offs, including potential negative transfer or degradation on typical speech benchmarks. A controlled comparison is therefore necessary to isolate the true contribution of speaker-independent warm-starting.

The \textbf{main contributions} of this work are as follows:

\begin{itemize}
    \item We studied whether speaker-independent adaptation provides a better initialization for speaker-level personalization.
    \item We proposed a speaker-disjoint two-stage adaptation framework that isolates the effect of initialization under identical per-speaker training conditions.
    \item We demonstrated consistent personalization gains across aphasic and dysarthric non-normative speech domains while analyzing the associated out-of-domain (OOD) trade-offs using two typical-speech benchmarks.
\end{itemize}

\begin{figure*}[t]
  \centering
  \includegraphics[width=\linewidth]{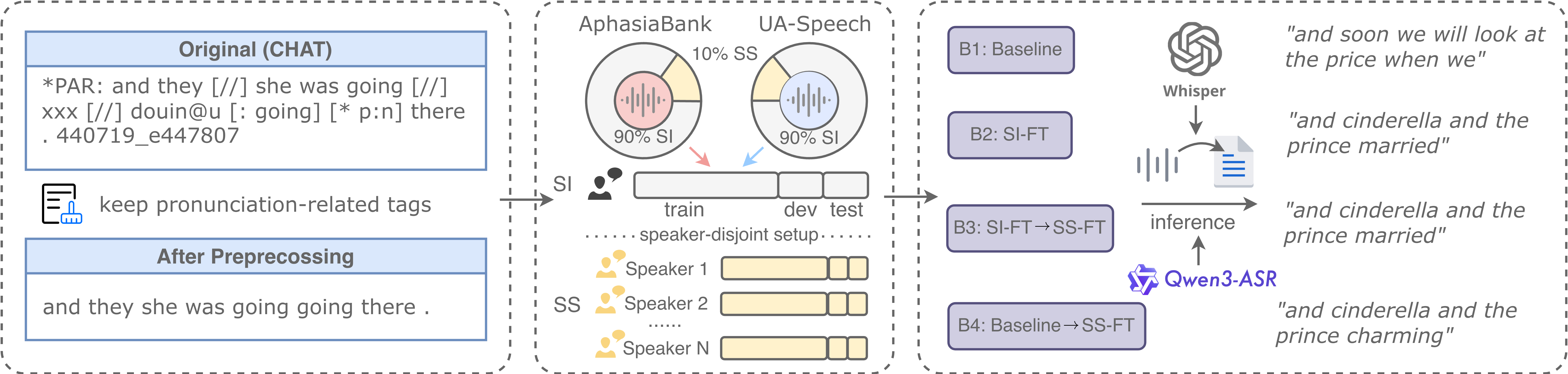}
  \caption{Overview of our methodology. Left: transcript normalization and filtering for AphasiaBank.
Middle: speaker-level partition into speaker-independent (SI) and speaker-specific (SS) groups (top 10\% speakers selected for personalization), with per-speaker train/validation/test splits.
Right: comparison of four experimental conditions (B1–B4), isolating the effect of speaker-independent initialization for speaker-specific adaptation.}
  \label{fig:system_overview}
\end{figure*}

\section{Methodology}
\label{sec:method}
Fig.~\ref{fig:system_overview} provides an overview of the proposed methodology, which consists of two main components: (i) two-stage adaption framework and comparison of four adaptation settings (B1–B4) in section~\ref{subsec:two-stage}, and (ii) dataset construction in section~\ref{subsec:Dataset}. 

\subsection{Two-stage Adaptation Framework}
\label{subsec:two-stage}
We propose a two-stage adaptation framework for non-normative ASR that separates population-level and speaker-level adaptation. The framework consists of two sequential steps:
\begin{itemize}
    \item \textbf{Stage 1 (Speaker-independent fine-tuning, SI-FT)} adapts a pre-trained ASR model using multi-speaker non-normative speech and produces a population-level model whose parameters are shifted toward the general non-normative speech patterns.
    \item \textbf{Stage 2 (Speaker-specific fine-tuning, SS-FT)} further adapts the model using individual speaker-level data to obtain a personalized model. 
\end{itemize}
This design allows us to examine whether population-level adaptation serves as a beneficial initialization prior for subsequent speaker-level personalization.
To isolate the effect of initialization, four experimental setups are compared:

\begin{itemize}
    \item \textbf{B1 (Baseline):} the vanilla ASR model without any adaptation.
    \item \textbf{B2 (SI-FT):} only SI-FT is performed, producing a population-level adapted model.
    \item \textbf{B3 (SI-FT$\rightarrow$SS-FT):} SI-FT followed by SS-FT for each target speaker, producing a set of personalized models. This corresponds to the proposed two-stage adaptation setting.
    \item \textbf{B4 (Baseline$\rightarrow$SS-FT):} direct SS-FT from the pre-trained baseline, using exactly the same per-speaker data and training setup as B3.
\end{itemize}
By comparing B3 and B4 under identical per-speaker training conditions, we isolate the contribution of speaker-independent initialization.

To evaluate whether this effect generalizes across architectures, experiments are conducted using two large-scale ASR models: Whisper~\cite{radford2023robust} and Qwen3-ASR~\cite{shi2026qwen3}.
\textbf{Whisper} is a Transformer-based encoder-decoder model trained on large-scale, diverse speech-to-text data with multi-task objectives, and it demonstrates strong zero-shot robustness across a range of benchmarks. 
\textbf{Qwen3-ASR} is a newly open-sourced all-in-one ASR model family. It is built on large-scale speech training data and post-trained from the multi-modal Qwen3-Omni foundation model~\cite{xu2025qwen3}. The models demonstrate strong audio understanding for speech recognition and language identification, and are designed for robust real-world speech recognition.

\subsection{Dataset Construction}
\label{subsec:Dataset}
\subsubsection{Datasets}
We employed two non-normative speech datasets, AphasiaBank~\cite{macwhinney2011aphasiabank} and UA-Speech~\cite{kim2008dysarthric}, and two typical-speech benchmarks, FLEURS~\cite{conneau2023fleurs} and TED-LIUM v3~\cite{hernandez2018ted}.
AphasiaBank is used to evaluate two-stage adaptation on conversational aphasic speech, while UA-Speech is used to verify findings on dysarthric speech.
TED-LIUM v3 and FLEURS (English) are used as OOD benchmarks to monitor potential degradation on typical speech after adaptation using non-normative speech datasets and are not included in adaptation training.

\textbf{AphasiaBank.}
AphasiaBank contains conversational recordings from speakers with aphasia with time-aligned transcripts in the CHAT format. It often reflects a mixture of impairments, including both language-level disruptions and articulatory deviations. In practice, speech from individuals with aphasia may exhibit dysarthria-like characteristics in addition to lexical or syntactic abnormalities, resulting in more heterogeneous speech patterns. This makes AphasiaBank a comparatively more complex and mixed-condition dataset.
In this work, we use only the protocol-compliant English subset, which comprises 525 speakers and approximately 449 hours of audio, and segment it into utterance-level clips based on the provided timestamps.

\textbf{UA-Speech.}
UA-Speech is an English dysarthric speech corpus with word-level utterances and intelligibility annotations, therefore does not require CHAT-style transcript processing.
In total, 16 dysarthric speakers are available, with approximately 66 hours of speech.
Microphone channel M5 is used in this work.

\textbf{Typical-speech benchmarks.}
The test sets of TED-LIUM v3 and FLEURS (English) are used only for evaluation and are not included in fine-tuning.
They provide OOD checks covering long-form public speech (TED-LIUM) and shorter, more diverse utterances (FLEURS).

\subsubsection{Transcript Normalization and Filtering}
The two non-normative datasets differ in annotation format and utterance granularity, which leads to different preprocessing requirements.
AphasiaBank provides conversational transcripts in the CHAT format, containing timestamps, linguistic annotations, retracing markers, and non-speech event symbols. These annotations are not ASR targets and may introduce label noise if used directly for training. In contrast, UA-Speech provides clean word-level references.

\textbf{Transcript normalization.}
For AphasiaBank, deterministic transcript normalization is applied to remove CHAT markers (e.g., retracing/repetition markers) to remove non-lexical annotations.
When explicit replacement annotations are present, a produced surface form is linked to an intended word or phrase, and the surface form is replaced with the intended word.
This includes cases where the surface form is a non-standard transcription (e.g., Unibet) or a word that is close in pronunciation to the intended target.
The same normalization rules are applied across all training conditions.
In contrast, UA-Speech provides single-word references and does not require this processing.

\textbf{SI-FT filtering.}
Errors in aphasia-related speech may reflect both articulatory deviations and language planning issues, including semantic errors.
To focus SI-FT on pronunciation adaptation, CHAT markers are used to select utterances that contain pronunciation-related tags \texttt{[* p: ...]} and to exclude utterances with semantic-error tags \texttt{[* s: ...]}.
This filtering reduces label noise caused by uncertain audio--text alignment and concentrates training on pronunciation-related variation. 
After filtering, 456 speakers remain, with approximately 7.67 hours of training data.

\subsubsection{Speaker-Disjoint Design}
To isolate the effect of speaker-independent initialization, we construct a speaker-disjoint experimental setup consisting of two non-overlapping groups of speakers: (i) a speaker-independent (SI) group used for population-level fine-tuning, and (ii) a speaker-specific (SS) group used for personalization.

\textbf{Speaker-level split.}
For each dataset, speakers are first partitioned at the speaker level into two non-overlapping groups.
The top 10\% speakers ranked by available duration are selected as target speakers for speaker-specific fine-tuning (SS-FT). All utterances from each selected speaker are used.
In AphasiaBank, this corresponds to 46 speakers.
In UA-Speech, two speakers (F02 and M12) are selected.
The per-speaker duration ranges from 1h33m to 7h.
All remaining speakers are assigned exclusively to the SI group.
Importantly, there is no speaker overlap between SI-FT and SS-FT.
This speaker-disjoint design ensures that population-level adaptation does not directly observe any target speaker used for personalization.

\textbf{Utterance-level split.}
For each target speaker in the SS group, all available utterances are split into train/validation/test subsets.
In AphasiaBank, an 8:1:1 ratio is used.
In UA-Speech, following the standard block-based setup, Block 1 and Block 3 are used for training, with 10\% of the training data held out for validation, and Block 2 is used for testing.
The same per-speaker training data and split are used for both two-stage personalization (B3, SI$\rightarrow$SS) and direct personalization from the pre-trained baseline (B4, Baseline$\rightarrow$SS).
This guarantees that any performance difference between the two conditions arises solely from the initialization strategy.

\textbf{SI training data.}
The SI-FT stage is trained using utterances from all speakers in the SI group (i.e., all non-target speakers) with no data from SS target speakers included.



\section{Experiments}

In the experiments, the identical two-stage adaptation is applied to Whisper-Large-v3 and Qwen3-ASR-1.7B using AphasiaBank and UA-Speech, respectively, to investigate how the method is affected by different backbones and domains.
ASR performance is evaluated primarily using word error rate (WER), with character error rate (CER) additionally reported. 
SI-FT is trained on multi-speaker data to obtain a population-level model, while SS-FT is trained separately for each target speaker to obtain personalized models.
To improve stability under limited per-speaker data, SS-FT freezes the lower half of the encoder layers and uses fewer training epochs.
Model selection is performed by choosing the checkpoint with the lowest validation WER.
SI-FT is trained on two NVIDIA H100 GPUs. For SS-FT, we train a separate model per target speaker using a single NVIDIA H100 GPU.

To minimize decoding-side confounds, Whisper models are decoded with the official Whisper implementation, while Qwen3-ASR is transcribed using its official pipeline. We align the task setup and decoding parameters for English transcription as closely as possible across the two frameworks.
Before computing WER, hypotheses and references are normalized using Whisper's English normalizer with a small set of additional rules (e.g., ``hafta'' $\rightarrow$ ``have to'') to reduce formatting-related mismatches.
Deterministic decoding is used (temperature=0).
Beam search with beam\_size=5 is applied to mitigate occasional degenerate repetition observed in a small number of samples. 



\section{Results and Discussion}
Table~\ref{tab:results} summarizes WER and CER across AphasiaBank and UA-Speech datasets and four adaptation settings (B1--B4), together with OOD evaluation on typical-speech benchmarks (FLEURS and TED-LIUM v3).

\subsection {Speaker-independent Adaptation}
We analyze on the effect of SI-FT by comparing the pre-trained baseline (B1) with SI-adapted models (B2), isolating the impact of population-level adaptation prior to personalization.

\textbf{In-domain non-normative-speech recognition.}
On AphasiaBank, the vanilla Whisper baseline (B1) yields a WER of 39.97 and a CER of 31.18 on the SI-FT test set, indicating a substantial mismatch to aphasic speech. After SI-FT (B2), performance improves to 27.70 WER and 21.97 CER.
Qwen3-ASR shows a similar pattern: the baseline achieves 34.07 WER and 26.26 CER, which improves to 27.68 WER and 21.83 CER after SI-FT.
A similar trend is observed on UA-Speech. For Whisper, SI-FT (B2) reduces WER from 113.34 to 37.08 and CER from 70.36 to 32.40. For Qwen3-ASR, performance improves from 78.38 WER / 59.10 CER to 34.81 WER / 31.06 CER.
These results confirm that population-level adaptation substantially reduces domain mismatch on non-normative speech.

\textbf{OOD impact analysis.}
On AphasiaBank, for Whisper, SI-FT (B2) slightly degrades FLEURS (WER 4.17$\rightarrow$4.74), but improving TED-LIUM v3 (WER 4.00$\rightarrow$2.87, a 28.3\% reduction), indicating that typical-speech performance is largely preserved, with even gains on long-form typical speech.
In contrast, although Qwen3-ASR achieves stronger baseline OOD performance (3.22 vs.\ 4.17 WER; 2.31 vs.\ 4.00 WER), adaptation degrades both FLEURS and TED-LIUM v3 (FLEURS 3.22 $\rightarrow$ 4.55; TED-LIUM v3 2.31 $\rightarrow$ 5.21), bringing its OOD performance closer to adapted Whisper. Overall, adaptation to conversational aphasic speech introduces limited and manageable OOD regression.

In contrast, UA-Speech SI-FT adaptation leads to noticeable OOD degradation on both benchmarks. 
For Whisper, FLEURS increases from 4.17 to 6.31 WER and TED-LIUM v3 from 4.00 to 4.56 WER, indicating a typical-speech trade-off when adapting to dysarthric speech. 
Qwen3-ASR exhibits higher OOD degradation, with improved performance on the dysarthric speech dataset coinciding with increased error rates on typical speech as FLEURS WER rises from 3.22 to 8.28 and TED-LIUM v3 WER rises from 2.31 to 14.61.

\definecolor{AphBG}{RGB}{255,225,225}   
\definecolor{UABG}{RGB}{225,235,255}    

\begin{table}[!htb]
    \caption{WER and CER for SI-FT and SS-FT on non-normative AphasiaBank and UA-Speech datasets, and OOD evaluation on typical-speech benchmarks (FLEURS and TED-LIUM v3) under four adaptation settings (B1, B2, B3, B4).}
    \label{tab:results}
    \centering
    \setlength{\tabcolsep}{3pt}
    \textbf{(a) Whisper-Large-v3}\par\smallskip
    \resizebox{\columnwidth}{!}{
        \begin{tabular}{l cc c cc c cc c cc}
            \toprule
            \multirow{2}{*}{Setup} &
            \multicolumn{2}{c}{\textbf{B1}} & &
            \multicolumn{2}{c}{\textbf{B2}} & &
            \multicolumn{2}{c}{\textbf{B3}} & &
            \multicolumn{2}{c}{\textbf{B4}} \\
            \cmidrule(lr){2-3}
            \cmidrule(lr){5-6}
            \cmidrule(lr){8-9}
            \cmidrule(lr){11-12}
            & \multicolumn{1}{c}{WER} & \multicolumn{1}{c}{CER} & &
              \multicolumn{1}{c}{WER} & \multicolumn{1}{c}{CER} & &
              \multicolumn{1}{c}{WER} & \multicolumn{1}{c}{CER} & &
              \multicolumn{1}{c}{WER} & \multicolumn{1}{c}{CER} \\
    
            \hline
            \multicolumn{12}{c}{\cellcolor{AphBG}\textbf{AphasiaBank}} \\
            SI-FT       & 39.97 & 31.18 & & \textbf{27.70} & \textbf{21.97} & & -- & -- & & -- & -- \\
            SS-FT       & 29.39 & 23.47 & & 23.90 & 19.59 & & \textbf{18.05} & \textbf{14.58} & & 20.11 & 16.34 \\
            \hline
            FLEURS      & \textbf{4.17} & \textbf{1.85} & & 4.74 & 2.24 & & 5.20 & 2.56 & & 4.70 & 2.25 \\
            TED-LIUM v3 & 4.00 & 2.60 & & \textbf{2.87} & \textbf{1.71} & & 3.14 & 1.91 & & 3.08 & 1.80 \\
            \hline
            
            \multicolumn{12}{c}{\cellcolor{UABG}\textbf{UA-Speech}} \\
            SI-FT       & 113.34 & 70.36 & & \textbf{37.08} & \textbf{32.40} & & -- & -- & & -- & -- \\
            SS-FT       & 153.31 & 91.90 & & 78.60 & 70.82 & & \textbf{58.17} & \textbf{57.76} & & 67.70 & 59.00 \\
            \hline
            FLEURS      & \textbf{4.17} & \textbf{1.85} & & 6.31 & 2.89 & & 7.92 & 4.02 & & 4.44 & 2.06 \\
            TED-LIUM v3 & \textbf{4.00} & \textbf{2.60} & & 4.56 & 3.06 & & 4.92 & 3.36 & & 4.22 & 2.81 \\
    
            \bottomrule
        \end{tabular}
    }

    \vspace{6pt}

    \textbf{(b) Qwen3-ASR-1.7B}\par\smallskip
    \resizebox{\columnwidth}{!}{
        \begin{tabular}{l cc c cc c cc c cc}
            \toprule
            \multirow{2}{*}{Setup} &
            \multicolumn{2}{c}{\textbf{B1}} & &
            \multicolumn{2}{c}{\textbf{B2}} & &
            \multicolumn{2}{c}{\textbf{B3}} & &
            \multicolumn{2}{c}{\textbf{B4}} \\
            \cmidrule(lr){2-3}
            \cmidrule(lr){5-6}
            \cmidrule(lr){8-9}
            \cmidrule(lr){11-12}
            & \multicolumn{1}{c}{WER} & \multicolumn{1}{c}{CER} & &
              \multicolumn{1}{c}{WER} & \multicolumn{1}{c}{CER} & &
              \multicolumn{1}{c}{WER} & \multicolumn{1}{c}{CER} & &
              \multicolumn{1}{c}{WER} & \multicolumn{1}{c}{CER} \\
    
            \hline
            \multicolumn{12}{c}{\cellcolor{AphBG}\textbf{AphasiaBank}} \\
            SI-FT       & 34.07 & 26.26 & & \textbf{27.68} & \textbf{21.83} & & -- & -- & & -- & -- \\
            SS-FT       & 23.04 & 18.27 & & 20.28 & 16.60 & & \textbf{18.80} & \textbf{15.44} & & 23.99 & 19.15 \\
            \hline
            FLEURS      & \textbf{3.22} & \textbf{1.41} & & 4.55 & 2.39 & & 5.03 & 2.82 & & 3.25 & 1.44 \\
            TED-LIUM v3 & \textbf{2.31} & \textbf{1.21} & & 5.21 & 3.96 & & 6.75 & 5.50 & & 2.55 & 1.43 \\
            \hline
            
            \multicolumn{12}{c}{\cellcolor{UABG}\textbf{UA-Speech}} \\
            SI-FT       & 78.38 & 59.10 & & \textbf{34.81} & \textbf{31.06} & & -- & -- & & -- & -- \\
            SS-FT       & 119.46 & 86.16 & & 72.37 & 71.04 & & \textbf{56.61} & \textbf{58.36} & & 100.19 & 84.28 \\
            \hline
            FLEURS      & \textbf{3.22} & \textbf{1.41} & & 8.28 & 5.68 & & 8.31 & 5.80 & & 6.35 & 4.23 \\
            TED-LIUM v3 & \textbf{2.31} & \textbf{1.21} & & 14.61 & 13.08 & & 13.24 & 11.64 & & 2.78 & 1.59 \\
            
            \bottomrule
        \end{tabular}
    }
    
\end{table}

\subsection{Speaker-Specific Personalization Warm-start}

We evaluate whether speaker-independent initialization improves downstream speaker-specific personalization by comparing two-stage adaptation (B3) with direct SS-FT from the pre-trained baseline (B4) under identical per-speaker conditions.
Across both datasets and backbones, B3 consistently outperforms B4 on the SS-FT evaluation sets.

\textbf{SS-FT results.} 
On AphasiaBank under SS-FT, B3 outperforms B4 on both models, achieving lower WER/CER on Whisper (18.05 vs.\ 20.11 WER; 14.58 vs.\ 16.34 CER) and on Qwen3-ASR (18.80 vs.\ 23.99 WER; 15.44 vs.\ 19.15 CER). 
On UA-Speech under SS-FT, the same trend largely holds: B3 improves WER over B4 on Whisper (58.17 vs.\ 67.70) and substantially on Qwen3-ASR (56.61 vs.\ 100.19), supporting SI-FT as a useful prior for downstream personalization. 

\begin{figure}[t]
  \centering
  \begin{subfigure}[t]{\columnwidth}
    \centering
    \includegraphics[width=\linewidth]
    {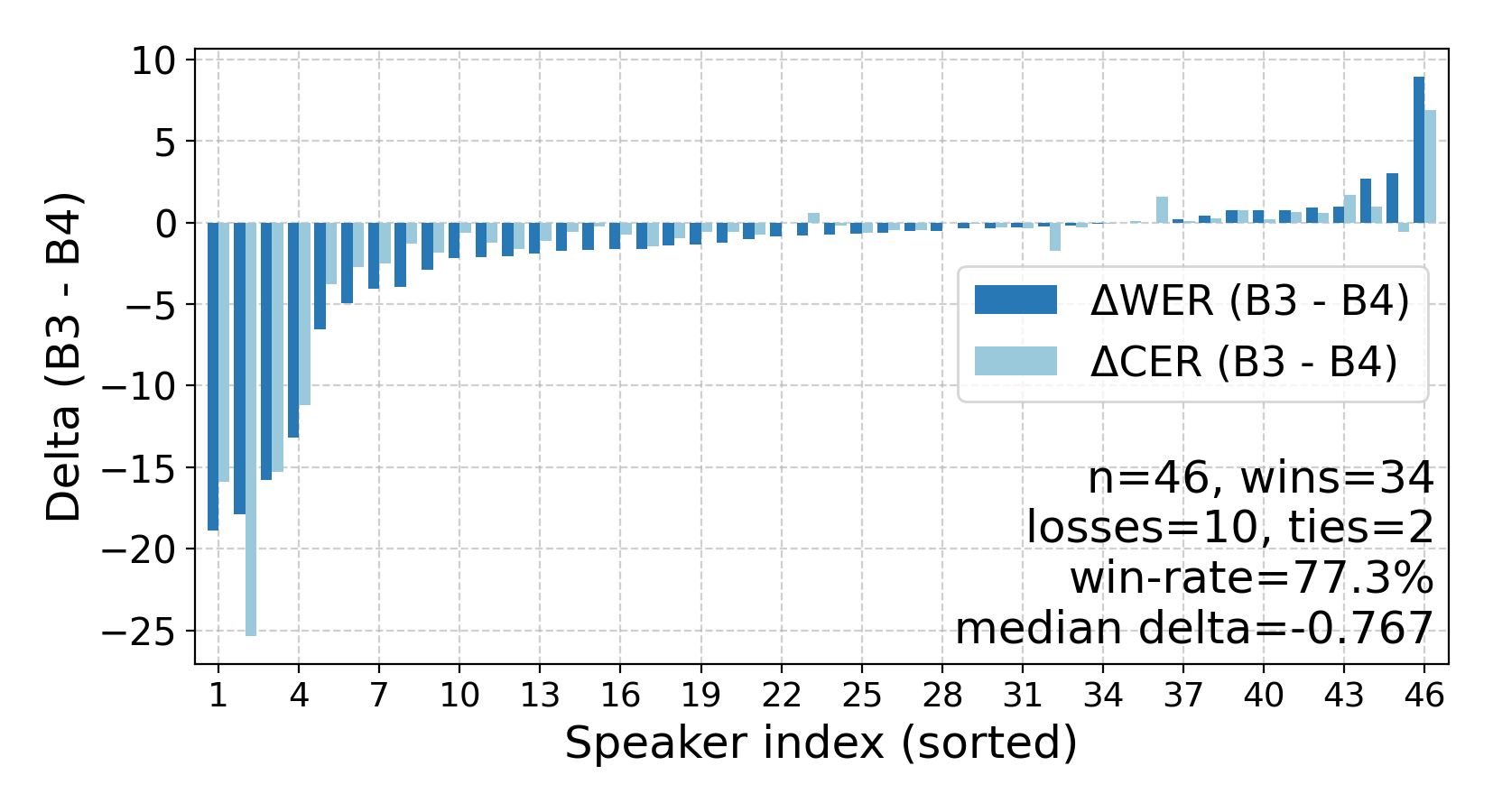}
    \caption{Whisper-Large-v3}
    \label{fig:delta_wer_b3_b4_whisper}
  \end{subfigure}
  \hfill
  \begin{subfigure}[t]{\columnwidth}
    \centering
    \includegraphics[width=\linewidth]
    {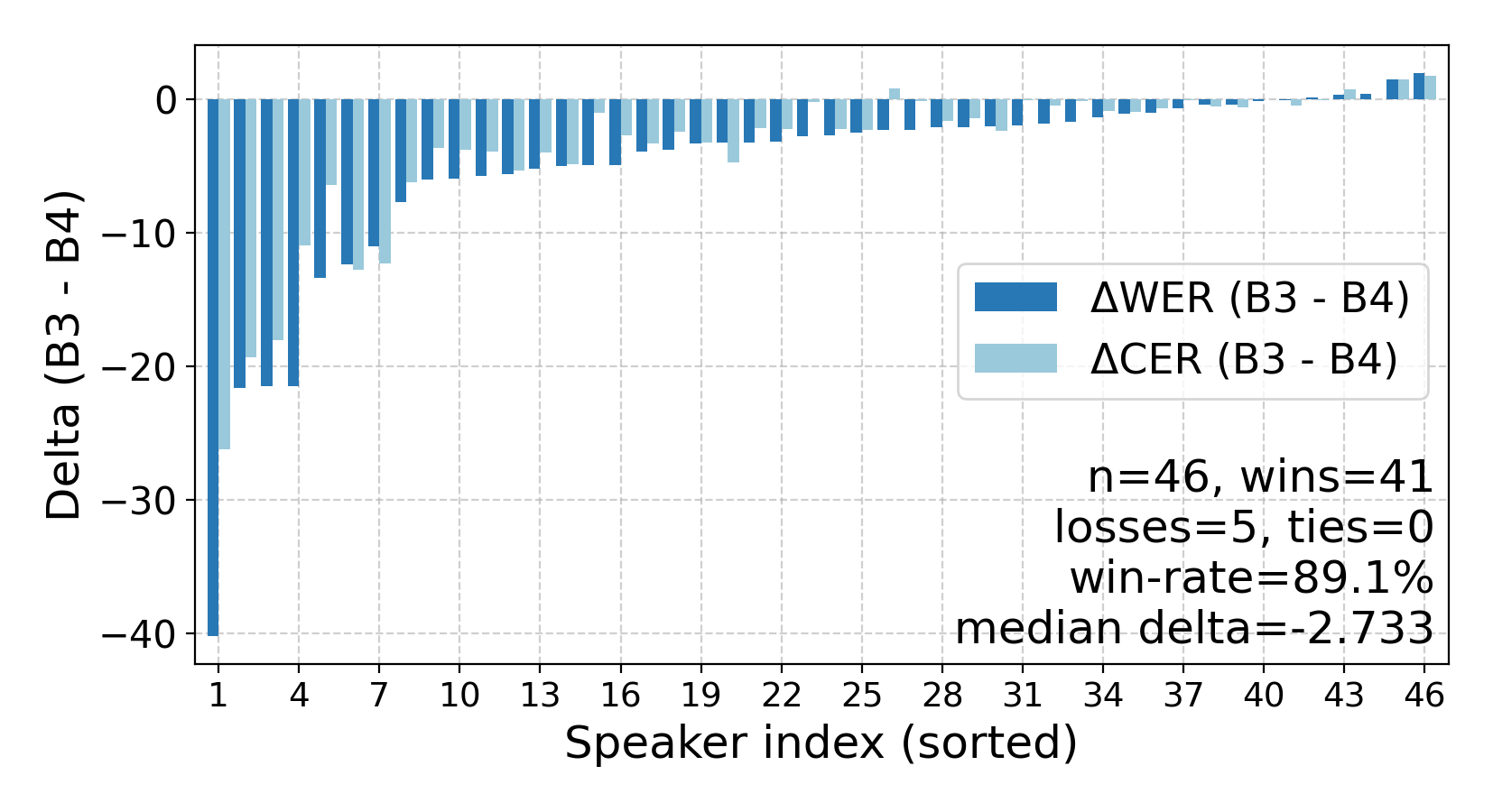}
    \caption{Qwen3-ASR-1.7B}
    \label{fig:delta_wer_b3_b4_qwen}
  \end{subfigure}

  \caption{Per-speaker $\Delta$WER (B3-B4) and $\Delta$CER (B3-B4) on AphasiaBank for two ASR systems.}
  \label{fig:delta_wer_b3_b4}
\end{figure}

\textbf{Speaker-level effects.}
Fig.~\ref{fig:delta_wer_b3_b4} further analyzes speaker-level effects on AphasiaBank by plotting per-speaker differences $\Delta$WER $=$ WER$_{B3}$ $-$ WER$_{B4}$.
For Whisper, 34 of 44 speakers (77.3\%, excluding ties) benefit from two-stage initialization with a median $\Delta$WER of $-0.767$. For Qwen3-ASR, improvements are observed for 41 of 46 speakers (89.1\%, excluding ties), with a median $\Delta$WER of $-2.733$. 
These results indicate that the warm-start advantage is broadly distributed rather than driven by a small subset of speakers.

On UA-Speech ($n=2$), both target speakers favor B3 over B4, with $\Delta$WER values of $-11.28$ and $-7.78$ on Whisper and $-49.81$ and $-37.35$ on Qwen3-ASR. This corresponds to a win-rate of 1.00 and median $\Delta$WER values of $-9.53$ (Whisper) and $-43.58$ (Qwen3-ASR).
Although limited in sample size, the improvement aligns with the AphasiaBank findings.

\textbf{OOD impact analysis.}
After adapting on AphasiaBank, OOD performance on typical-speech benchmarks remains largely stable, indicating that personalization gains can be achieved with minimal regression on typical speech. 
In contrast, UA-Speech fine-tuning exhibits a clearer OOD trade-off, both typical-speech benchmarks degrade after adaptation. This suggests that population-level adaptation to dysarthric speech may shift the model distribution more aggressively, increasing mismatch under typical-speech test conditions. One possible explanation is that UA-Speech contains more word-level, highly non-normative realizations than the sentence-level conversational errors in AphasiaBank, inducing a distribution shift and, consequently, a more pronounced trade-off.



\section{Conclusion and Future Work}

In this work, we revisited non-normative ASR personalization from the perspective of initialization and asked whether population-level SI adaptation can serve as a stronger prior than a generic foundation model for subsequent personalization.
Across AphasiaBank and UA-Speech, the proposed two-stage framework consistently outperforms direct personalization under identical per-speaker training conditions.
These results suggest that population-level adaptation on non-normative speech effectively reduces the distribution gap at personalization, leading to more stable and sample-efficient adaptation.
We further analyzed the OOD effects of non-normative adaptation on typical-speech benchmarks.
Adaptation on AphasiaBank largely preserves typical-speech accuracy, whereas UA-Speech adaptation exhibits a clearer trade-off, indicating that the degree of distribution shift and domain characteristics can materially affect generalization.
Overall, our findings support SI adaptation as a practical warm-start strategy for non-normative ASR personalization, while emphasizing the need to manage typical-speech degradation when adapting to highly divergent domains.

Future work will explore cross-dataset warm-starting (e.g., SI-FT on one non-normative corpus followed by SS-FT on another), more stable personalization via PEFT methods (e.g., adapters/LoRA), and regularization strategies to better preserve typical-speech performance.


\section{Generative AI Use Disclosure}
Generative AI tools were used in a limited and supportive capacity during this work. Specifically, such tools assisted in polishing the manuscript and the codebase for data pre-processing and visualization. 
All generated content, including code and written material, was carefully reviewed, validated, and edited by the authors. The authors take full responsibility for the accuracy, originality, and integrity of the final manuscript.

\section{Acknowledgments}
This work was supported by the Dutch Research Council (NWO) through the VENI Talent Programme under Grant No. 21828.

\bibliographystyle{IEEEtran}
\bibliography{mybib}

\end{document}